\documentclass[aps,prl,twocolumn,,groupedaddress,showpacs,nofootinbib]{revtex4}
\usepackage{epsfig}
\usepackage[dvipsnames,usenames]{color}

\begin{document}

\title{Magnetic transition in a correlated band insulator}

\author{A. Euverte$^1$, 
S. Chiesa$^2$, R.T.~Scalettar$^3$, and
G.G. Batrouni$^{1,4}$}

\affiliation{$^1$INLN, Universit\'e de Nice-Sophia Antipolis, CNRS;
1361 route des Lucioles, 06560 Valbonne, France }

\affiliation{$^2$Department of Physics, College of William \& Mary,
Williamsburg, VA 23185, USA}

\affiliation{$^3$Physics Department, University of California, Davis,
California 95616, USA}

\affiliation{$^4$Institut Universitaire de France}

\begin{abstract}
The effect of on-site electron-electron repulsion $U$ in a band
insulator is explored for a bilayer Hubbard Hamiltonian with opposite
sign hopping in the two sheets.  The ground state phase diagram is
determined at half-filling in the plane of $U$ and the interplanar
hybridization $V$ through a computation of the antiferromagnetic (AF)
structure factor, local moments, single particle and spin wave spectra,
and spin correlations.  Unlike the case of the ionic Hubbard model, no
evidence is found for a metallic phase intervening between the Mott and
band insulators.  Instead, upon increase of $U$  at large $V$, the
behavior of the local moments and of single-particle spectra give
quantitative evidence of a crossover to a {\color{black}Mott} insulator state
preceeding the onset of magnetic order.  Our conclusions generalize
those of single-site dynamical mean field theory, and show that
including interlayer correlations results in an increase of the single
particle gap with $U$.
 \end{abstract}  

\maketitle

\noindent
\underbar{Introduction:} Whether interactions might drive metallic
behavior in two dimensional disordered systems, where disorder just
marginally succeeds in localizing all the eigenstates, is a question
that has been the subject of considerable experimental and theoretical
scrutiny\cite{lee85,belitz94,dobrosavljevic12}.  It is natural to ask
the same question concerning band insulators which likewise have
vanishing $dc$ conductivity in the absence of interactions.  In the
case of band insulators, carrier density plays an especially central
role, since the band must be precisely filled.  This lends an
additional complexity to the issue, since interactions might also give
rise to Mott insulating and magnetic behavior.

The possible connection between disordered interacting systems, and
correlated band insulators is made more concrete by considering the
Anderson model, where random site energies couple to the local density,
and the `ionic' Hubbard model (IHM)\cite{hubbard81} which has a
superlattice potential where the site energy has a regular structure,
taking two distinct values on the sublattices of a bipartite lattice.
On the one hand, it is plausible that the same physical effects that
could cause a metallic transition for random site energies, the
reduction of charge inhomogeneity and resulting delocalization of the
electronic wave functions by interparticle repulsion, would also be
operative in the patterned case.  On the other hand, momentum is still a
good quantum number in the presence of a regular array of site energies,
suggesting possible differences between the effect of $U$ in the two
situations.

The approximations made in the most simple, single site, Dynamical Mean
Field Theory (DMFT) approach\cite{georges96} to the treatment of electron-electron
interaction emphasize some of the possible nuances in attempting to
elucidate the physics of correlated band insulators.  Single site DMFT
can capture the band insulator (and how it differs from an Anderson
insulator) by incorporating a density of states with $N(E_{\rm F})=0$.
However, it also minimizes the role of momentum, and hence blurs some of
the distinction between band and Anderson insulators.

DMFT has, in fact, been used to explore whether correlations can drive a
band insulator metallic.  Garg {\it et al.}~found\cite{garg06} that for
the IHM, treated within single site DMFT, the band gap becomes zero at a
critical $U_{c1}$, with a Mott gap re-emerging at a larger $U_{c2}$.  In
between, the system is metallic.  A subsequent cluster DMFT study of
Kancharla {\it et al.}\cite{kancharla07} which incorporated
antiferromagnetic correlations, found a phase diagram with somewhat
different topology, but still exhibiting an intermediate region which
was suggested to have bond ordered wave character.

\noindent
\underbar{Model:}
In this paper,
we shall consider a bilayer Hubbard Hamiltonian:
\begin{eqnarray}
\label{Hamiltonian}
{\hat\mathcal H}&=& -\sum_{\langle {\bf jk} \rangle, l, \sigma } 
t_{l}^{\phantom{\dagger}} 
(c_{{\bf j},l,\sigma}^\dagger c_{{\bf k},l,\sigma}^{\phantom{\dagger}} + 
c_{{\bf k},l,\sigma}^\dagger c_{{\bf j},l,\sigma}^{\phantom{\dagger}} ) 
\nonumber \\
&&-V \sum_{{\bf j}, \sigma}(c_{{\bf j},1,\sigma}^\dagger 
c_{{\bf j},2,\sigma}^{\phantom{\dagger}} 
+ 
c_{{\bf j},2,\sigma}^\dagger c_{{\bf j},1,\sigma}^{\phantom{\dagger}} ) 
- \sum_{{\bf j},l,\sigma} \mu_l^{\phantom{\dagger}}
n_{{\bf j},l,\sigma}^{\phantom{\dagger}} 
\nonumber \\ 
&& + U \sum_{{\bf j},l}(n_{{\bf j},l,\uparrow}^{\phantom{\dagger}}-\frac{1}{2})
(n_{{\bf j},l,\downarrow}^{\phantom{\dagger}}- \frac{1}{2})
\label{eq:hamilt}
\end{eqnarray}
which provides a specific realization of the effect of electronic
correlation in band insulators.  In Eq.~\ref{eq:hamilt}, $c_{{\bf
j},l,\sigma}^\dagger (c_{{\bf j},l,\sigma}^{\phantom{\dagger}}) $ are
creation(destruction) operators for fermions of spin $\sigma$ on site
${\bf j}$ of layer $l=1,2$.  The intralayer hoppings are
$t_{l}^{\phantom{\dagger}}$ between near-neighbor sites ${\bf j,k}$ of a
two dimensional square lattice, and the interlayer hopping is $V$.
Correlation is introduced in the model through the on-site repulsion
$U$.  We have included a chemical potential $\mu_l$ for generality.
However, here we focus on the half-filled case $\mu_l=0$.

If the in-plane hoppings are chosen with opposite sign,
$t_1=-t_2\equiv t$, Eq.~\ref{eq:hamilt} is a band insulator at $U=0$
with $E_{\bf q}= \pm \sqrt{\epsilon_{\bf q}^2 + V^2}$.  
(In the remainder of this paper $t=1$ will be taken
as the unit of energy.) This bears a
strong resemblance to the dispersion relation of the 2D IHM whose
superlattice potential $ \Delta \sum_{\bf j} (-1)^{\bf j} n_{\bf j}$
similarly has the effect of opening a band gap at ${\bf q}=(\pi,\pi)$,
altering the $\Delta=0$ dispersion relation $\epsilon_{\bf q}$ to
$E_{\bf q}= \pm \sqrt{\epsilon_{\bf q}^2 + \Delta^2}$.  Real space
Quantum Monte Carlo (QMC)\cite{paris07} has supplemented DMFT studies of
the IHM\cite{garg06} by allowing for magnetic order and concluded that
$U$ could cause the appearance of a metallic phase.  However, despite
the similarity in $E_{\bf q}$ between the IHM and Eq.~\ref{eq:hamilt},
the physics of the two models is fundamentally different:  the bilayer
has twice as many allowed ${\bf q}$ points and uniform average density,
$\langle n_{{\bf j},l,\sigma}\rangle = \frac{1}{2}$.  This is in
contrast to the staggered charge density wave pattern in the presence of
the superlattice potential in the IHM.  This difference, combined with
the local character of the interaction, is at the origin of the contrast
in the ground state properties which we will present.

As with the IHM, the Hamiltonian in Eq.~\ref{eq:hamilt} has been
previously studied within the DMFT formalism \cite{sentef09}, with a
model hybridizing two bands with identical, semi-elliptical density of
states.  DMFT finds a scenario remarkably similar to the one observed
for the metal-insulator case in the standard single band model at
half-filling: a first order transition between band and Mott insulator
characterized by a discontinuous change in the double
occupancy.  Remarkably, upon increasing the interaction,
DMFT also predicts that the single-particle gap should monotonically
shrink in stark contrast with the behavior in the Mott phase where the
gap grows monotonically with $U$.

We will explore these issues using determinant Quantum Monte Carlo
(DQMC) \cite{blankenbecler81}.  This method allows an exact
calculation\cite{trotter} of the properties associated to the
Hamiltonian in Eq.~\ref{eq:hamilt}, on lattices of finite spatial
extent.  Here we will show results for systems consisting of two sheets
of up to $N=14\times14$ sites.

\begin{figure}[ht]
\centerline{\epsfig{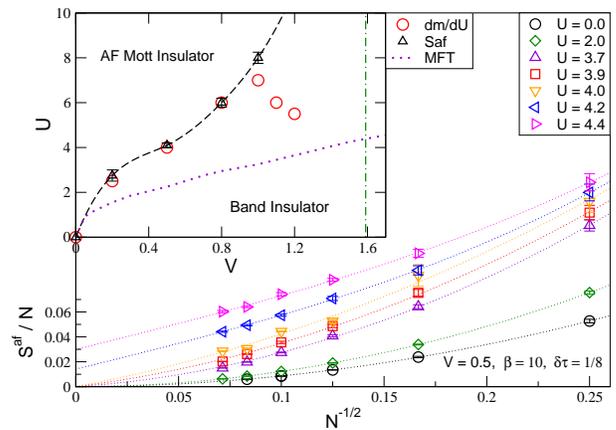}}
\caption{Finite-size scaling of the AF structure factor at $V=0.5$.
  Symbols are DQMC data for $S^{\rm af}$.  Lines are fits to
  third-order polynomials in the inverse linear lattice size
  $1/\sqrt{N}$. The inset shows the transition line computed with
  DQMC (dashed) and MFT (dotted). The vertical line corresponds
  to the critical $V$, predicted by studies on the Heisenberg model,
  above which no magnetic long order is possible. Circles correspond to
  maxima in $dm/dU$ as a function of $U$ at constant $V$ (see Fig.~\ref{m_U}).}
\label{Saf_L_V0.5} 
\end{figure}

\noindent
\underbar{Magnetic transition:} We start our discussion of the phases
of Eq.~\ref{eq:hamilt} by looking for possible antiferromagnetic
long-range order (LRO).  The most direct signature is the
thermodynamic extrapolation of the in-layer structure factor
(which has the same value on the two layers),
\begin{eqnarray}
S({\bf q}) &=& \frac{1}{6N} \sum_{{\bf j,k},l} \, \langle \,
\sigma^z_{{\bf k},l} \sigma^z_{{\bf j},l} + 2\,\sigma^-_{{\bf k},l}
\sigma^+_{{\bf j},l} \, \rangle \, e^{i {\bf q} \cdot ({\bf k} - {\bf
    j})} \\ \sigma^z_{{\bf j}} &=& c_{{\bf j}\uparrow}^\dagger c_{{\bf
    j}\uparrow}^{\phantom{\dagger}} -c_{{\bf j}\downarrow}^\dagger
c_{{\bf j}\downarrow}^{\phantom{\dagger}}; \phantom{aa} \sigma^+_{{\bf
    j}} = c_{{\bf j}\uparrow}^\dagger c_{{\bf
    j}\downarrow}^{\phantom{\dagger}}; \phantom{aa} \sigma^-_{{\bf j}}
= c_{{\bf j}\downarrow}^\dagger c_{{\bf
    j}\uparrow}^{\phantom{\dagger}} \nonumber
\label{eq:saf}
\end{eqnarray}
converging to a non-zero value as $N \rightarrow \infty$.
We will focus on
antiferromagnetism, $S^{\rm af}=S(\pi,\pi)$, which is the expected
dominant magnetic instability at half-filling.  Because of the
continuous spin symmetry and the fact that we are in two dimensions,
we expect LRO only at $T=0$.

The finite size scaling analysis necessary to locate the critical value
of $U$ for onset of magnetic order is presented in Fig.~\ref{Saf_L_V0.5}
for $V=0.5$.  At weak coupling $U \lesssim 4.0$ there is no LRO.  As the
on-site repulsion increases, LRO sets in around $U \approx 4.2$.  Our
value of $U_c$ is significantly smaller than the DMFT estimate of
approximately 5.5 \cite{sentef09} which was however computed for a
transition to a  Mott insulating state without long range magnetic
order.

We repeated the finite size scaling analysis for several other values of
$V$ and obtained the phase diagram shown in the inset of Fig.1. There is
a qualitative difference between the DQMC results and the transition
line mean field theory (MFT) predicts. Beyond $V=0.5$ the DQMC curves
rises much more sharply than the MFT one as a result of the competition
between interplanar singlet formation and AFM correlation which is
lacking in the mean-field description. Note also that known results for
the Heisenberg bilayer\cite{hida92,wang06} imply that for $V >
V_c=1.59$, no order is established regardless of the magnitude of $U$.
Determining the values of $U_c$ as $V_c$ is approached becomes quickly
intractable for $V>1.0$ since the energy scale at which magnetic
correlations develop decreases and fluctuations in the DQMC measurements
of long range correlations increase. 

\begin{figure}[ht]
\centerline{\epsfig{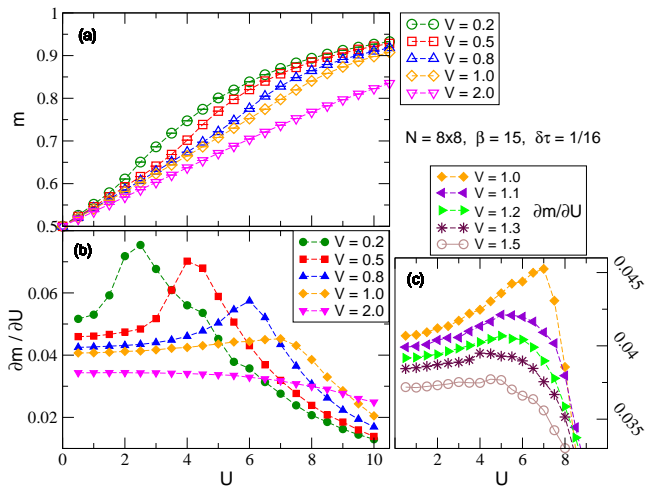}}
\caption{ (a) Local moment  $m$ 
as a function of $U$ for $8\times8$ layers at $\beta=15$ (symbols).  (b)
first derivative of the local moment with respect to $U$, $\partial m
/ \partial U$, shows a peak at the transition to the Mott insulating
state.
(c) A close up view of (b) shows the peak is no longer present for any
$U$ above $V_c \approx 1.2$. }
\label{m_U} 
\end{figure}

\vskip0.10in
\noindent
\underbar{Local moments:} Within DMFT\cite{sentef09}, the phase
boundary is determined by a discontinuity in the double occupancy $d$,
the latter being related to the local moment $m$ by,
\begin{eqnarray*}
 m =\frac{1}{N}\sum_{\bf j} \langle (\sigma^z_{\bf j})^2\rangle 
= 1 - \frac{2}{N}\sum_{\bf j} \langle n_{{\bf j}\,\uparrow} n_{{\bf j}\,\downarrow}\rangle = 1-2d.
\end{eqnarray*}
Local moment formation is the key signature for the onset of Mott
insulating behavior and it has been previously reported to happen
discontinuously at a Mott metal-insulator transition within other
approaches as well, such as path-integral renormalization
group~\cite{kashima01} and variational Monte Carlo~\cite{tahara08}.
Figure~\ref{m_U}a shows the dependence of $m$ on interaction strength
for several values of $V$ with no evidence of the sharp discontinuity
found in DMFT. 

However, for small $V$ ({\em e.g.} $V=0.5$), we found that the magnetic
transition is located in correspondence to a maximum in ${\partial
m}/{\partial U}$ (see Fig.~\ref{m_U}b).  Numerical differentiation does
not allow us to establish whether there is an actual singularity in the
behavior of ${\partial m}/{\partial U}$ --- so that one could use this
quantity to characterize a phase transition --- or whether the maximum
is a simple manifestation of a cross over to the local moment regime.
At larger $V$ the value of $U$ where the maximum appears is reduced
(Fig.~\ref{m_U}c and circles in Fig.1), presumably as a consequence of
the increased electron localization on the interplane bonds, whereas
$U_c$ for the onset of AFLRO is expected to grow monotonically.  This
decoupling of the behavior of local moments from magnetism is suggestive
of the possibility of an intervening Mott insulating state with no
broken symmetries.

\begin{figure}[ht]
\centerline{\epsfig{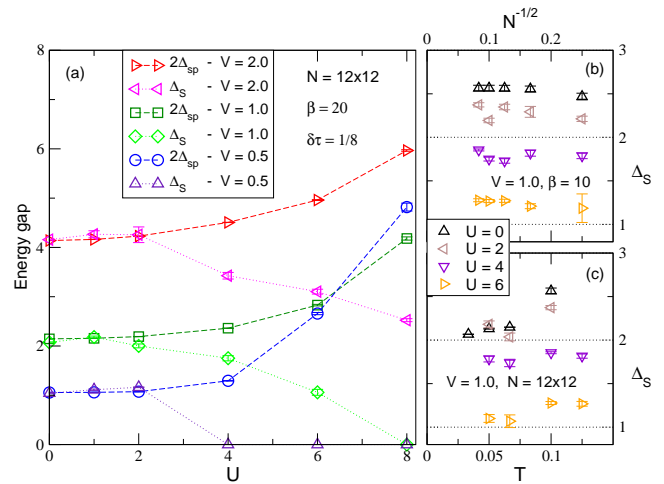}}
\caption{Main panel: Gaps in the single particle spectral function and
  the spin-spin correlation function are shown as functions of
  interaction $U$ for several values of interlayer hopping $V$.  At
  weak $U$, the spin gap $\Delta_S$ is precisely twice the single
  particle gap $\Delta_{\rm sp}$, as expected in a Fermi liquid phase.
  The smaller panels at right show the finite size (top) and finite
  temperature (bottom) effects for the spin gap.
\label{gaps} }
\end{figure}

\vskip0.10in
\noindent
\underbar{Energy gaps and spectral functions:} We now investigate the
evolution of the energy spectra. The single-particle gap $\Delta_{sp}$
and spin excitation gap $\Delta_{S}$ were extracted from the imaginary
time-dependent correlations
\begin{eqnarray}
G(\tau) &=& \sum_{{\bf i},\sigma}\langle 
c^{\phantom{\dagger}}_{{\bf i},\sigma}(\tau)
c^{\dagger}_{{\bf i},\sigma}(0)  \rangle \propto e^{-\tau\Delta_{sp}}
\nonumber \\
\chi(\tau) &=& \sum_{{\bf i},\sigma}\langle \sigma_{{\bf i},\sigma}^z(\tau)
\sigma_{{\bf i},\sigma}^z(0)  \rangle \propto e^{-\tau\Delta_S}.
\nonumber 
\end{eqnarray}
Figure~\ref{gaps}a shows the evolution of these gaps when the
interaction $U$ is increased, for different values of the interlayer
hybridization $V$.  Starting from the non-interacting limit, where the
single-particle gap and the spin gap are expected to be respectively
equal to $V$ and $2V$, the two quantities follow opposite evolutions
regardless of whether there is a tendency toward AFM (small $V$) or
singlet formation (large $V$).  In particular, we found that for all
three values of $V$ considered in Fig.~3, the effect of correlation is
negligible up to $U\simeq 2$, in agreement with the findings of Sentef
\cite{sentef09}.  However, we observe a significant discrepancy
between the effect of correlation in DMFT and in DQMC: in DQMC the
single particle gap $\Delta_{\rm sp}$ shows no indication of the
shrinking trend predicted by DMFT, not even in the small $U$ limit
where the role of long range correlation should be negligible and the
paramagnetic solution is likely the correct ground state within
single-site DMFT.  We checked that the values of the gaps are
converged in both size of the cluster (Fig.~\ref{gaps}b) and
temperature (Fig.~\ref{gaps}c).  Although it is certainly the case
that this difference in physics is associated with the effects of non
local spatial correlations on spectral properties, as noted in
discussions of cluster extensions of
DMFT\cite{maier05,haferman09,katanin09}, our calculation does not
allow to address the interesting issue of whether the discrepancy
originates from the neglect of interlayer singlet correlation or
intralayer short-range anti-ferromagnetism. The former represents a
somewhat more severe failure of single-site DMFT as the difference
would be largely independent of the underlying lattice structure and
on whether or not the latter supports any ordering tendencies.

\begin{figure}[ht]
\centerline{\epsfig{figure=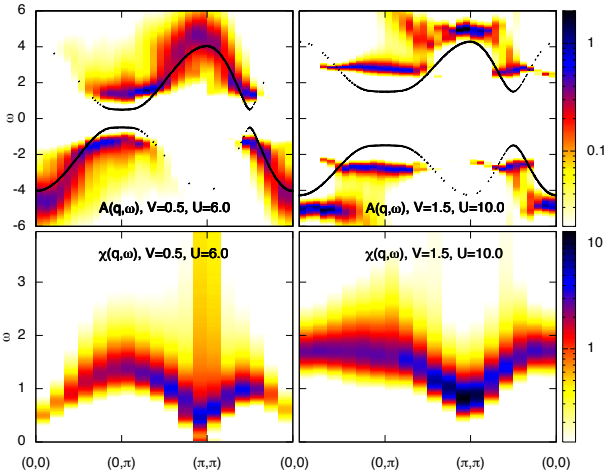,width=8cm,angle=0}}
\caption{ Top row: single particle spectral function in the presence(left)
and absence (right) of AFM order. Bottom row: same as top row
but for the spin spectral function. Results are computed on a $N=12\times12$ 
cluster at $\beta=20$. Lines in the top panels are the corresponding energy
bands when $U=0$.
\label{Sz_wk} }
\end{figure}

In-layer momentum-resolved single-particle and spin 
excitation spectra ($A({\bf q},\omega)$ and $\chi({\bf q},\omega)$) 
are obtained by inverting the integral equations
\begin{eqnarray}
G({\bf q},\tau) &=&  
\int_0^\beta A({\bf q}, \omega)
\frac{e^{-\omega \tau}}{1 + e^{-\beta \omega}}
d\omega
\nonumber \\
\chi({\bf q},\tau) &=&
\int_0^\beta \chi({\bf q}, \omega)
\frac{e^{-\omega \tau}}{1 - e^{-\beta \omega} }
d\omega. \nonumber
\end{eqnarray}
using the maximum entropy method \cite{gubernatis91,beach04}.  $G({\bf
q},\tau)$ and $\chi({\bf q},\tau)$ are the in-layer momentum resolved
counterparts of the correlation functions previously introduced.
Figure~\ref{Sz_wk} shows the  single-particle (top panels) and spin
(bottom panels) spectral densities and compare an AFM situation (left)
against a case where no order is found by a finite size scaling
analysis(right).  Thanks to the Goldstone theorem, the spin spectra
provide a complementary indication of the presence or lack of AFM long
range order. We verified that, indeed, parameter regimes that were
predicted to be AFM by scaling analysis are characterized by the
existence of a massless mode at $(\pi,\pi)$ which is conspicuously
absent in paramagnetic cases. The single particle spectrum, on the other
hand, helps in characterizing the paramagnetic phase more precisely as
it shows 
{\color{black} an almost rigid shift of the non-interacting bands, a
behavior indicative of a Mott insulating regime.}
At large $V$, but still below $V_c$, our results therefore
suggest that, upon increase of $U$, the system shows a first crossover
to a  {\color{black}featureless Mott} insulating state and then a transition into the
anti-ferromagnet. 
{\color{black} It is the crossover that can be most directly contrasted with 
the DMFT scenario which predicts a split narrowing resonance in the 
weak-to-intermediate $U$ range and then a transition to a Mott insulator.}

\begin{figure}[ht]
\centerline{\epsfig{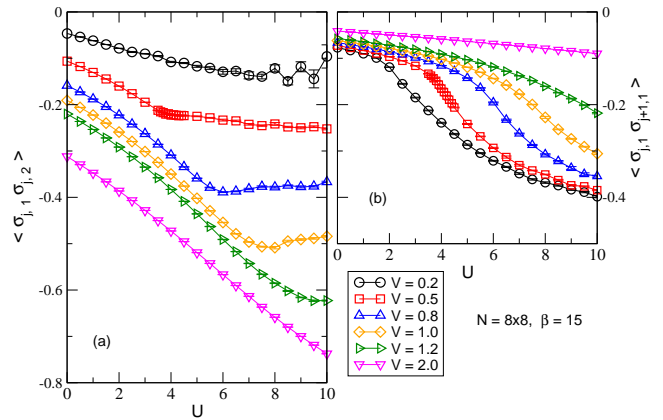}}
\caption{ Near-neighbor real space spin-spin correlations (a) across
  the layers and (b) within an individual plane.  At small $V<1.2$ the
  correlations converge to finite values characterizing the magnetic
  ordered phase, while for large $V=2.0$, the system is made of almost
  decorrelated singlets.  }
\label{SaXZ} 
\end{figure}

\noindent
\underbar{Spin correlations:} The real space spin correlations across
the layers $ \langle \sigma_{{\bf j} 1} \cdot \sigma_{{\bf j} 2} \rangle$
are shown in Fig.~\ref{SaXZ}a.  The generic behavior of bilayer models
(and related Hamiltonians like the periodic Anderson model) is the
development of singlets with increasing $V$ at fixed $U$, and the
associated destruction of AFLRO, signalled by a growth in $ \langle
\sigma_{{\bf j} 1} \cdot \sigma_{{\bf j} 2} \rangle$.  The development of
such interplane spin correlations can be seen in Fig.~\ref{SaXZ} by
comparing the different curves at fixed $U$. The evolution at fixed
$V$ also
provides consistent indications of the underlying physics previously
inferred from the structure factor $S^{\rm af}$, the local moment $m$,
and the excitation gaps.  Specifically, the interlayer spin
correlations first increase as interactions are turned on, but then
have a kink, or even turn over, as the AF phase is entered.  For
$V=0.5$ for example, the kink appears at $U \approx 4.0$.

The intra-plane real space nearest-neighbor spin correlations are shown
in panel (b) of Fig.~\ref{SaXZ}.  They increase monotonically (in
absolute value) with $U$ for all $V$, indicating that the on-site
Hubbard $U$ enhances short range intraplane
antiferromagnetism\cite{foot2}.  This quantity offers yet another local
diagnostic for the onset of order as it shows an inflection point in
close correspondence to the transition. Moreover, comparison of the
results in the two panels at $V=0.5$ and small $U$ reveals that it is
the interplane spin correlation that grows more rapidly.  We take this
as a further indication that the discrepancy in the behavior of the
single particle gap between our calculation and single site DMFT is not
due to the presence of intralayer short-ranged magnetic order but to the
inclusion, by the DQMC approach used here, of interlayer singlet
correlations.

\vskip0.25in

\noindent
\underbar{Conclusions:} In this paper we studied the effect of
introducing local interaction in the band insulator formed by a bilayer
with opposite sign of the hopping integral.  We found strikingly
different physics from the ionic Hubbard model owing to the fact that
the system is perfectly homogeneous and accompanied by a tendency toward
singlet formation as the band gap increases.  As the strength of the
interaction is increased, and below a critical interplane hybridization,
a transition to a Mott insulator with antiferromagnetic order ensues.
This transition was studied by examining several physical observables
such as the magnetic structure factor, the local moments,
single-particle and spin excitations resolved in both energy and
momentum, and spin correlations. The behavior of $\partial m /\partial
U$ {\color{black} and spectral functions} suggests that, as $V$ grows, 
the magnetic transition is preceeded by a cross-over into a featureless 
{\color{black}Mott} insulating state.
A more subtle question is 
whether such cross-over may, in fact, be a transition.  Obviously, this 
is a delicate point that requires validation from calculations on larger 
clusters and the use of a direct estimator for $\partial m/\partial U$ 
rather than the finite difference employed in this work.

\noindent
\underbar{Acknowledgements:}
We thank F. Assaad for invaluable help with the analytic continuation
code; and Z.~Bai, A.~Tomas, and J.~Perez for their work optimizing DQMC.
This work was supported by: the CNRS-UC Davis EPOCAL LIA joint
research grant; the NSF grant OISE-0952300; the ARO Grant 56693-PH;
and ARO Award W911NF0710576 with funds from the DARPA OLE Program. 



\begin{thebibliography}{99}
\bibitem{trotter}
The only approximation is the introduction of a finite discretization 
of the inverse temperature
$\beta=1/T = L \Delta \tau$.
The resulting `Trotter errors' can be eliminated through
an extrapolation to $\Delta \tau=0$.  For the $\Delta \tau$ values
chosen here, we have verified that
such an extrapolation does not shift the physical
results significantly.
\bibitem{foot2}
For fixed non-zero $T$, the spin correlations would ultimately
go down as $U$ increases owing to the $1/U$ dependence of $J$.
In Fig.~\ref{SaXZ}, however, since $\beta=15$ we are sufficiently close
to the ground state not to observe that effect.
\end{thebibliography}
\end{document}